\documentclass[11pt]{article}

\usepackage[final]{acl}
\usepackage{booktabs}
\usepackage{tabularx}
\usepackage{multirow}
\usepackage{graphicx}
\usepackage{svg}
\usepackage{times}
\usepackage{latexsym}
\usepackage{amsmath}  
\usepackage{amssymb}  
\usepackage{listings}         
\usepackage{xcolor}           
\usepackage[T1]{fontenc}

\usepackage[utf8]{inputenc}

\usepackage{microtype}

\usepackage{inconsolata}

\usepackage{graphicx}
\usepackage[framemethod=default]{mdframed}
\usepackage{multicol}
\newmdenv[
  linewidth=0.8pt,
  linecolor=black,
  backgroundcolor=white,
  skipabove=0pt,
  skipbelow=0pt,
  innerleftmargin=10pt,
  innerrightmargin=10pt,
  innertopmargin=8pt,
  innerbottommargin=8pt,
  frametitlefont=\bfseries,
  frametitlerule=true,
  frametitlerulewidth=0.8pt,
  frametitlebackgroundcolor=gray!10
]{appendixframe}

\title{ResLPO: Residual Listwise Preference Optimization for Long-Context Review Ranking}

\author{Hao Jiang \\
  Nanyang Technological University\\
  \texttt{jianghao907@gmail.com} \\\And
  Zhi Yang \\
  Peking University\\
  \texttt{zhiyang25@stu.pku.edu.cn} \\
  \And 
Annan Wang \\
  Nanyang Technological University\\
  \texttt{annan001@e.ntu.edu.sg} \\ \AND
  Yichi Zhang  \\
  Independent Researcher \\
  \texttt{yichizhang0926@gmail.com} \\
  \And
Weisi Lin\thanks{Corresponding author} \\
  Nanyang Technological University\\
  \texttt{WSLin@ntu.edu.sg}
  }

\begin{document}

\maketitle
\begin{abstract}

Review ranking is pivotal in e-commerce for prioritizing diagnostic and authentic feedback from the deluge of user-generated content. While large language models have improved semantic assessment, existing ranking paradigms face a persistent trade-off in long-context settings. Pointwise scoring is efficient but often fails to account for list-level interactions, leading to miscalibrated top-$k$ rankings. Listwise approaches can leverage global context, yet they are computationally expensive and become unstable as candidate lists grow.
To address this, we propose Residual Listwise Preference Optimization (ResLPO), which formulates ranking as listwise representation-level residual correction over a strong pointwise LLM scorer. ResLPO first produces calibrated pointwise scores and item representations, then applies a lightweight encoder over the representations to predict listwise score residuals, avoiding full token-level listwise processing. 
We also introduce a large-scale benchmark for long-context review ranking with human verification. Experiments show ResLPO improves NDCG@k over strong pointwise and listwise baselines and remains robust as list length increases.

\end{abstract}

\section{Introduction}
\begin{figure}[htbp]
  \centering
  \includegraphics[width=0.98\linewidth]{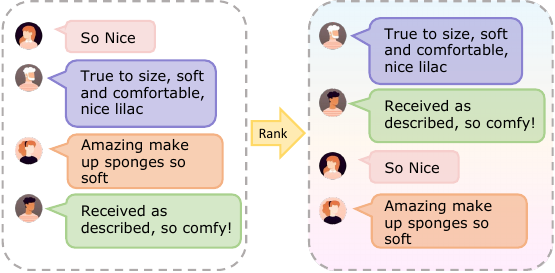}
  \caption{Illustration of review ranking for the product ``Retro Lazy Suede Half Slippers.'' The left panel shows randomly ordered reviews, while the right panel shows the ranked list. The top-ranked review is more informative and detailed than the second, whereas the bottom review is ranked last because it is unrelated to the product.The full review texts are provided in Appendix~\ref{app2}.}
  \label{versus}
\end{figure}
The modern e-commerce ecosystem is predicated not merely on the exchange of goods, but on the exchange of information \cite{ai2017learning,bi2020transformer,yan2022personalized}. User-generated reviews have become the primary mechanism for trust verification and product discovery \cite{hou2024bridging}. However, the exponential growth of online feedback has created a paradox of choice: a popular product may accumulate tens of thousands of reviews, rendering the vast majority invisible. As seen in Fig.\ref{versus}, the utility of a review is not absolute but relative; a review is only valuable if it offers diagnostic information distinct from what the user has already read. Traditional ranking algorithms, often relying on simple metadata or recency, fail to parse the semantic nuance required to surface such content. Consequently, users are frequently forced to sift through redundant or irrelevant text, highlighting the urgent need for ranking systems that can intelligently curate truthful and informative content to the top of the list.

The emergence of Large Language Models (LLMs), such as Gemini\cite{team2023gemini} and GPT \cite{achiam2023gpt}, with their extensive world knowledge and reasoning capabilities, has fundamentally reshaped the landscape of ranking. Recent advancements have seen the deployment of LLMs in various ranking paradigms, yet each suffers from distinct limitations when applied to review ranking \cite{zhu2025large}. Pointwise methods \cite{liu2025llm4ranking, gera2025justrank,xu2025precisezeroshotpointwise,zhang2025leveragingreferencedocuments}, while straightforward and scalable, score documents in isolation. They suffer from a ``myopic'' view, estimating relevance probability without regard for list-level interactions such as redundancy \cite{liu2025llm4ranking}. For instance, a pointwise model might assign identical high scores to five high-quality reviews of a product, the model may struggle to induce a consistent ordering among these five items, and it may also fail to promote a sixth review that provides a different perspective. This calibration bias can lead to suboptimal top-$k$ results that degrade user experience.

Conversely, listwise ranking models~\cite{gupta2025scalable,liu2025lipo,cai2025k,wu2025context,Reddy2024FIRSTFI,zhao2024selfcalibratedlistwisereranking,liu2025corankingcollaborativeranking} are often viewed as the theoretical ideal because they can incorporate the global context of the candidate set. However, current LLM-based listwise rankers face substantial efficiency and stability challenges. In practice, adding a single new review may require re-processing the entire review list for the same product, leading to redundant computation. As the number of candidates grows, the input context length increases rapidly, making inference expensive due to the quadratic complexity of self-attention. Moreover, long-context listwise ranking can suffer from performance degradation and hallucinations, where the model under-attends to reviews in the middle of the context window or produces permutations not grounded in the input. Related pairwise~\cite{Qin2023LargeLM,liu2025harnessingpairwiseranking} and setwise approaches~\cite{chen2024softmax,wang2025autorule,Zhuang2023ASA} mitigate some issues, but their inference cost grows exponentially with the number of reviews. This creates a dilemma: one must choose between the efficiency of pointwise methods and the contextual awareness of listwise methods, with no existing framework effectively bridging the gap for long-context ranking.


To address these challenges, we introduce \textbf{Res}idual \textbf{L}istwise \textbf{P}reference \textbf{O}ptimization (\textbf{ResLPO}), a residual listwise correction framework that bridges pointwise scoring and list-level interactions without token-level listwise re-encoding.
Specifically, a fine-tuned LLM produces calibrated pointwise scores along with compact review representations, and a lightweight set encoder attends over the representation sequence to predict list-conditioned score residuals that correct ordering errors caused by redundancy and score compression.
This decoupling preserves the semantic strengths and scalability of pointwise scoring, while injecting global list awareness with substantially reduced computation compared to token-level listwise prompting.

A further obstacle to progress in review ranking is the absence of public, standardized benchmarks tailored to the review ranking setting. Although real-world products often have long review lists, existing public resources are typically designed for product-level retrieval or ranking and do not provide dense, listwise supervision for ordering reviews within the same item. This limitation hinders consistent comparison and systematic analysis of list-level behavior as candidate set size varies. To close this gap, we construct a large-scale benchmark from real-world e-commerce reviews with item-level candidate lists, dense ranking labels, and human verification, and we will release it publicly to support reproducible research.


Our contributions are summarized as follows:
\begin{itemize}
    \item We propose ResLPO, to our knowledge the first residual listwise preference optimization framework that bridges pointwise scalability and listwise global context for long-context review ranking, addressing the effectiveness--efficiency trade-off.
    \item We construct and will publicly release a large-scale review ranking benchmark derived from the Amazon Reviews 2023 dataset, with dense listwise supervision and human verification, filling a gap in domain-specific evaluation resources.
    \item Extensive experiments show that ResLPO achieves state-of-the-art ranking performance, remains robust as list length increases, and avoids the instability of generative listwise rankers under long contexts.
\end{itemize}

\section{Related Work}

Ranking has long been studied in information retrieval and recommendation. Before the recent wave of LLM-based rankers, mainstream approaches largely relied on unsupervised lexical matching and neural encoders that map queries and documents into comparable representations. More recently, LLMs ~\cite{team2023gemini,achiam2023gpt,liu2024deepseek,bai2023qwen}, have further advanced ranking by enabling stronger semantic reasoning and instruction following, ushering in a new era of generative ranking. Below we review these lines of work and position ResLPO.

\subsection{Unsupervised and Encoder-Based Ranking}

Early ranking methods rely on unsupervised lexical matching that scores documents using corpus-level token statistics. TF--IDF~\cite{ramos2003using} and BM25~\cite{robertson2009probabilistic} are representative examples, offering strong efficiency, scalability, and interpretability, but they largely model lexical overlap and often miss semantic relevance and nuanced utility signals required by review ranking. With the success of Transformer architectures~\cite{vaswani2017attention}, neural encoder-based rankers became a dominant paradigm by encoding queries and documents into dense representations and computing relevance via representation comparison~\cite{yu2025unbiased}, enabling semantic matching beyond exact term overlap. Similar encode-then-compare designs have also proven effective in other modalities such as vision transformers~\cite{han2022survey}. Nevertheless, encoder-based rankers can still struggle to capture fine-grained list-level interactions when candidate sets are large, and they may be less effective at modeling deeper semantic preferences needed for high-quality reranking.

\subsection{Pointwise LLM Ranking}

LLM-based ranking methods build on these foundations by leveraging the world knowledge and reasoning capabilities of LLMs, which have also driven recent advances in multi-turn reasoning \cite{ma2026tspo} and dynamic dialogue management \cite{hu2026context}. Pointwise methods score each candidate document independently, typically producing a relevance or utility score for a query--document pair. This paradigm is straightforward and scalable, and it naturally supports large candidate sets because inference is linear in the number of documents. Recent work studies pointwise prompting and training for LLM ranking and provides systematic evaluations~\cite{liu2025llm4ranking,gera2025justrank}. However, since candidates are assessed in isolation, pointwise ranking can be insensitive to list-level interactions (e.g., redundancy among top results), which may lead to calibration issues in the final top-$k$ list.

\subsection{Pairwise and Setwise LLM Ranking}

Pairwise methods compare two candidates at a time and infer a preference relation, then aggregate pairwise outcomes into a final ordering. Compared to pointwise scoring, pairwise comparison provides an explicit relative signal, but the required number of comparisons grows quickly with candidate set size, increasing inference cost. Setwise variants extend pairwise comparison by ranking or selecting within small groups, aiming to improve efficiency while preserving relative judgments. Recent studies explore such pairwise and setwise formulations and objectives for LLM ranking~\cite{chen2024softmax,wang2025autorule}, but scaling to long review lists still requires many comparisons and non-trivial aggregation.

\subsection{Listwise LLM Ranking}

Listwise methods condition on the entire candidate set and generate an ordered list directly, which is often viewed as the most context-aware paradigm. Recent work develops listwise objectives and strategies for LLM ranking~\cite{gupta2025scalable,liu2025lipo,cai2025k,wu2025context}. While listwise ranking can capture global context and inter-document dependencies, it can be expensive and unstable for long contexts, as the input grows with the number of candidates and token-level self-attention becomes costly.
Our work targets the gap between pointwise scalability and listwise awareness. We retain the efficiency of pointwise scoring, while introducing a lightweight residual mechanism that injects list-level context at the representation level, enabling global re-ordering without token-level listwise processing.


\begin{figure*}[t]
  \centering
  \includegraphics[width=\textwidth]{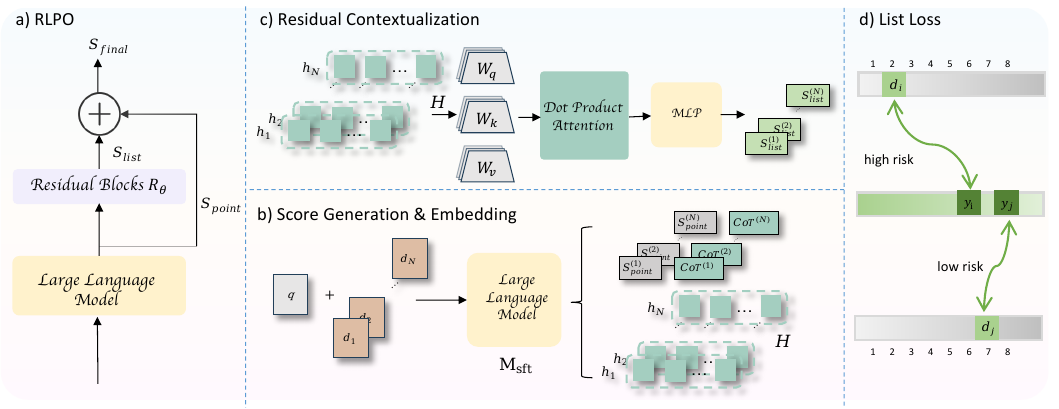}
  \caption{The overall framework of ResLPO}
  \label{fig:overall}
\end{figure*}

\section{Review Ranking Benchmark}
\label{sec:dataset}

To facilitate research on long-context review ranking, we construct a comprehensive benchmark derived from real-world e-commerce scenarios, which we will release publicly to support future work and reproducibility. In this section, we detail the data collection pipeline and the human verification protocol used to ensure label quality.

\begin{table}[h]
\centering
\resizebox{\columnwidth}{!}{%
\begin{tabular}{l|rr|rrr}
\hline
\textbf{Category} & \textbf{Products} & \textbf{Reviews} & \textbf{Avg. Revs} & \textbf{Avg. Len} & \textbf{Avg. Score} \\ \hline
Baby Products & 1,119 & 76,371 & 68.3 & 39.4 & 7.04 \\
Fashion & 2,065 & 50,177 & 24.3 & 26.4 & 6.59 \\
Software & 348 & 99,872 & 287.0 & 24.3 & 5.65 \\
All Beauty & 1,935 & 98,292 & 50.8 & 36.7 & 6.67 \\ \hline
\textbf{Total / Avg.} & \textbf{5,467} & \textbf{324,712} & \textbf{59.4} & \textbf{32.0} & \textbf{6.43} \\ \hline
\end{tabular}%
}
\caption{Statistics of the constructed benchmark. }
\label{tab:dataset_stats}
\end{table}

\subsection{Data Collection and Annotation}

We source our data from the Amazon Reviews 2023 dataset~\cite{hou2024bridging}. To ensure domain diversity, we specifically select products from four distinct categories: \textit{All\_Beauty}, \textit{Fashion}, \textit{Baby\_Products}, and \textit{Software}. These categories represent a wide range of review characteristics.

To obtain high-quality ranking labels, we employ Gemini-2.5-Pro \cite{comanici2025gemini} as an expert annotator. As illustrated in Appendix.\ref{app1}, the model is prompted to evaluate each review based on a multi-dimensional schema, considering its intrinsic attributes (e.g., \emph{content richness}, \emph{usefulness}, and \emph{quality}) as well as its extrinsic relevance to the instruction $q$. Table~\ref{tab:dataset_stats} summarizes the statistics of the constructed benchmark. The dataset maintains a high density of reviews per product, providing a challenging testbed for listwise ranking models.

\subsection{Human Verification}
\label{sec:human_eval}

Since review utility can be subjective, we conduct a two-stage human evaluation to validate the reliability of the LLM-generated labels.

\begin{figure}[t]
  \centering
  \includegraphics[width=0.9\linewidth]{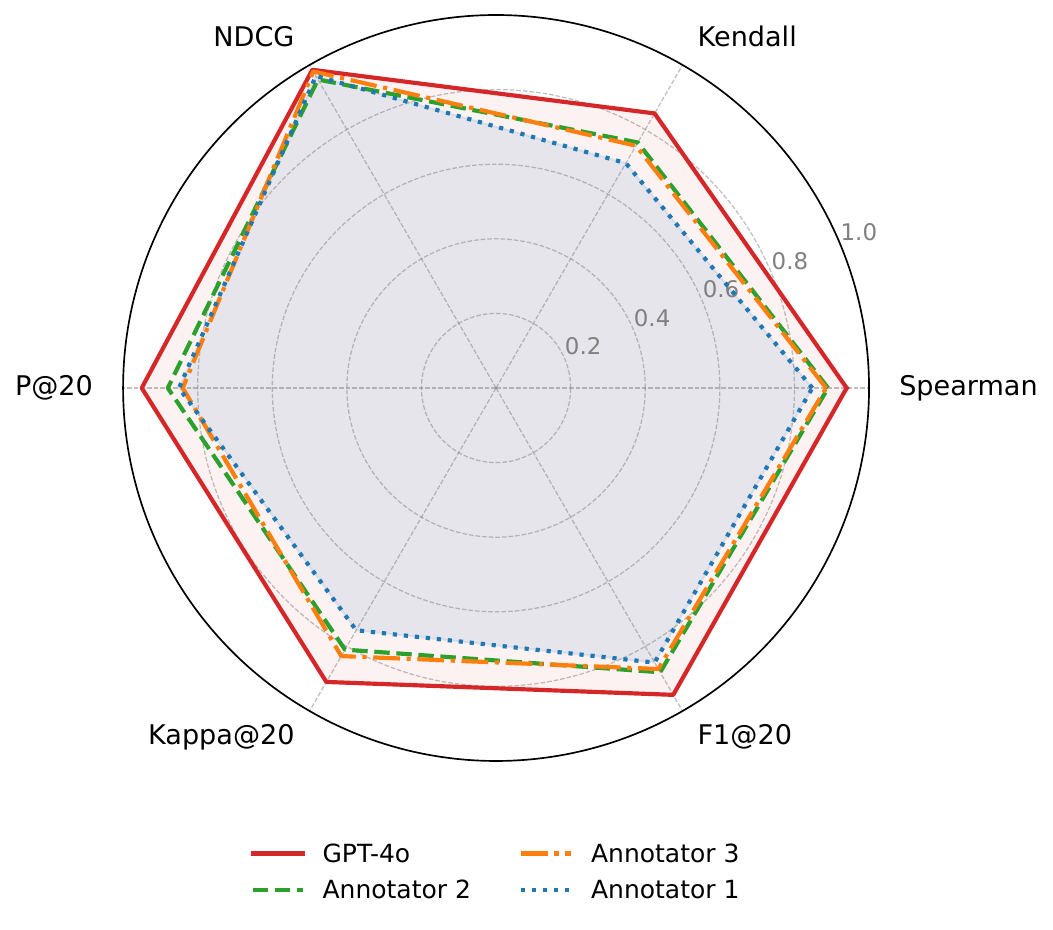}
  \caption{Performance comparison of human annotators and GPT-4o against the dataset ground truth. The radar chart depicts agreement across six metrics (e.g., NDCG, Spearman), highlighting the high quality and consistency of the generated labels.}
  \label{fig:radar_chart}
\end{figure}

\paragraph{Listwise Ranking Consistency.} 
First, we randomly sample a subset of products and their corresponding candidate reviews (up to 50 items per list, 1k reviews in total). We employ three human annotators and GPT-4o to independently rank these lists. As shown in Appendix.\ref{app3}, to mitigate cognitive load and ensure precision, annotators follow a bubble sort-inspired protocol: they perform iterative pairwise comparisons to establish a total ordering of the reviews. 
We assess annotation quality by measuring agreement between three human annotators, and GPT-4o as an additional reference, against our ground-truth rankings using rank correlation and top-$k$ consistency metrics. Figure~\ref{fig:radar_chart} shows consistently high agreement across annotators, with NDCG \cite{wang2013theoretical} ranging from 0.955 to 0.980, indicating strong consistency on listwise ordering. Correlation metrics are also stable, with Spearman \cite{essam2022comparison} ranging from 0.848 to 0.890 and Kendall ranging from 0.696 to 0.760. These results suggest that the LLM-generated labels largely align with human judgments despite the inherent subjectivity of review helpfulness.


\paragraph{Pairwise Accuracy Check.} 
Second, to further quantify label accuracy, we conduct a pairwise preference test. We randomly sample 2,000 review pairs from the same product and ask human experts to identify the more helpful review in each pair. The results demonstrate that our generated labels achieve a pairwise accuracy exceeding 90\%, confirming that the relative orderings in our benchmark are semantically sound and aligned with human preferences.

\section{ResLPO Framework}


In this section, we formally present the ResLPO framework. ResLPO is designed to resolve the dichotomy between the scalability of pointwise scoring and the contextual awareness of listwise ranking. We first detail the hybrid architecture, which disentangles ranking into intrinsic relevance estimation and global contextual correction. Subsequently, we derive our optimization objective, which aligns the residual gradient updates directly with the non-differentiable NDCG metric via a Lambda-weighted mechanism.

\subsection{The ResLPO Architecture}
As seen in Fig.~\ref{fig:overall}, the fundamental hypothesis of ResLPO is that the utility of a review $d_i$ given a query $q$ (i.e., a product-aware prompt that includes the product title and other available product metadata) can be decomposed into two orthogonal components: (1) Intrinsic Relevance, derived from the semantic alignment between $q$ and the review text, and (2) Contextual Utility, which captures the relative value of the review (e.g., diversity, redundancy) conditional on the candidate list $D=\{d_1,\dots,d_N\}$.

Pointwise strategies are myopic, estimating only the former. Listwise strategies attempt to model the joint distribution $P(D\mid q)$, but often succumb to the quadratic cost of token-level self-attention over long contexts, especially when the candidate set changes and the full list must be re-processed.
ResLPO adopts a parameter-efficient paradigm: a fully fine-tuned LLM backbone produces pointwise scores (with chain-of-thought (CoT) rationales) and compact document embeddings, while a lightweight, trainable Residual Head operates on the embedding sequence to regress a list-conditioned score adjustment for each review.

\subsubsection{Phase 1: Semantic Score Generation and Encoding (Pointwise)}
Let $\mathcal{M}_{\text{SFT}}$ denote a Large Language Model (e.g., Mistral-7B) after supervised fine-tuning (SFT) for review assessment. Given a candidate set $D=\{d_1,\dots,d_N\}$, we process each query--review pair $(q,d_i)$ independently. For each review $d_i$, $\mathcal{M}_{\text{SFT}}$ is trained to assess its intrinsic attributes (e.g., \emph{content richness}, \emph{usefulness}, and \emph{quality}) as well as its \emph{relevance} to the query $q$, and to generate a structured output consisting of a numerical pointwise score $s_{\text{point}}^{(i)}$ and a chain-of-thought (CoT) rationale. The CoT serves to strengthen semantic understanding and improve self-correction during generation. In addition to the generated score, we extract a compact semantic representation $\mathbf{h}_i \in \mathbb{R}^d$ from the last hideen layer. Formally,
\begin{equation}
\text{CoT}_i,\; s_{\text{point}}^{(i)},\; \mathbf{h}_i \;=\; \mathcal{M}_{\text{SFT}}(q, d_i).
\end{equation}

\subsubsection{Phase 2: Residual Contextualization (Listwise)}
To capture global dependencies, we introduce a lightweight Residual Self-Attention Block. As shown in Figure~\ref{fig:overall}(c), this module operates on the sequence of compressed review embeddings for a product,
$H = [\mathbf{h}_1, \dots, \mathbf{h}_N]$,
rather than on the token sequence of a single review. This design enables $N\times N$ interactions at the embedding level with low overhead.
We apply a standard multi-head self-attention (MHSA) layer to model inter-review relations:
\begin{equation}
    H_{\text{ctx}} = \text{LayerNorm}\bigl(H + \text{MHSA}(H)\bigr).
\end{equation}
Intuitively, MHSA serves as a comparison operator that can capture list-level effects such as redundancy (e.g., down-weighting a review that is semantically similar to others). We then project each context-aware representation to a scalar delta score:
\begin{equation}
    \Delta s_{\text{list}}^{(i)} = \text{MLP}\bigl(H_{\text{ctx}}^{(i)}\bigr),
\end{equation}
which represents a list-conditioned adjustment to the pointwise prior.
During residual contextualization, the backbone $\mathcal{M}_{\text{SFT}}$ is kept frozen, and we optimize only the parameters of the residual block, reducing training cost while preserving the capabilities learned during SFT.

\subsubsection{Score Aggregation}
Inspired by \cite{qiu2025gated,he2016deep}, the final ranking score $s_{\text{final}}^{(i)}$ is formulated as a residual correction:
\begin{equation}
    s_{\text{final}}^{(i)} = s_{\text{point}}^{(i)} + \alpha \cdot \Delta s_{\text{list}}^{(i)}
\end{equation}
where $\alpha$ is a learnable scaling factor (initialized to 0). This ResNet-style formulation provides a stable optimization landscape: the model starts by mimicking the pointwise ranker and gradually learns to perturb scores only when the global context necessitates a re-ordering.

\subsection{Importance-Aware Listwise Loss}
\label{sec:lambda_loss}

Ranking metrics such as NDCG are position-sensitive: errors near the top of the list incur much larger utility loss than those near the tail~\citep{wang2013theoretical}. To reflect this, we adopt an importance-aware objective that scales learning signals by the (approximate) NDCG change induced by correcting ordering mistakes, following the LambdaRank/LambdaLoss philosophy.

Given a query/product with candidate set $D=\{d_1,\dots,d_N\}$, let $y_i$ be the ground-truth utility label and $s_i \triangleq s_{\text{final}}^{(i)}$ be the predicted score. We define
\begin{equation}
\mathrm{gain}(y)=2^{y}-1,
\qquad
\mathrm{disc}(k)=\frac{1}{\log_2(k+1)}.
\end{equation}
Let $\pi^\star$ be the permutation that sorts labels in descending order (ties broken deterministically). The ideal discounted cumulative gain is
\begin{equation}
\mathrm{IDCG}=\sum_{k=1}^{N}\mathrm{gain}(y_{\pi^\star(k)})\,\mathrm{disc}(k).
\end{equation}

For any pair $(i,j)$ with $y_i>y_j$, we compute a non-negative importance weight based on the current predicted ranking $\pi$ induced by sorting scores $s$ (used only to obtain ranks). Let $r_i$ and $r_j$ denote their 1-indexed ranks under $\pi$. We define
\begin{equation}
\begin{aligned}
\Delta \mathrm{gain}_{ij} &= \mathrm{gain}(y_i)-\mathrm{gain}(y_j),\\
\Delta \mathrm{disc}_{ij} &= \mathrm{disc}(r_i)-\mathrm{disc}(r_j).
\end{aligned}
\end{equation}
and the associated NDCG change magnitude
\begin{equation}
\Delta_{ij}
=
\frac{1}{\mathrm{IDCG}}
\left|
\Delta \mathrm{gain}_{ij}\cdot \Delta \mathrm{disc}_{ij}
\right|.
\label{eq:delta_ndcg}
\end{equation}

We then optimize the NDCG-weighted pairwise logistic loss
\begin{equation}
\ell_{ij}
=
\log\!\bigl(1+\exp(-(s_i-s_j))\bigr),
\end{equation}
\begin{equation}
\mathcal{L}_{\text{ResLPO}}
=
\sum_{i=1}^{N}\sum_{j=1}^{N}
\mathbb{I}[y_i>y_j]\;
\Delta_{ij}\;
\ell_{ij}.
\label{eq:ResLPO_lambdaloss}
\end{equation}

The objective in Eq.~\eqref{eq:ResLPO_lambdaloss} is differentiable with respect to scores $s$. The only non-smooth operation is the sorting step used to compute ranks $(r_i,r_j)$ for $\Delta_{ij}$. In practice, we treat $\Delta_{ij}$ as a detached weight (i.e., no gradient flows through sorting), while gradients propagate through $\ell_{ij}$. We ignore pairs with $y_i=y_j$ and apply deterministic tie-breaking when computing $\pi^\star$ for IDCG.


\section{Experiment}
To rigorously validate the efficacy of Residual Listwise Preference Optimization (ResLPO) in the domain of long-context information retrieval, we conducted an exhaustive series of experiments. These experiments were designed not merely to demonstrate incremental improvements in ranking metrics, but to probe the fundamental capacity of Large Language Models (LLMs) to reason over extensive, noise-laden contexts when aligned via listwise objectives. 
Our investigation is structured around four primary research questions (RQs) that guide the subsequent analysis:
\begin{itemize}
    \item \textbf{RQ1 (Comparative Effectiveness):} To what extent does ResLPO outperform existing pairwise (e.g., DPO) and listwise (e.g., LiPO) alignment baselines in ranking high-utility reviews?
    \item \textbf{RQ2 (Long-Context Robustness):} How does performance change as the candidate list length increases, and does ResLPO mitigate long-context degradation?
    \item \textbf{RQ3 (Generalization Across Domains):} How well does ResLPO transfer across product categories with different review distributions?
    \item \textbf{RQ4 (Efficiency and Scalability):} What are the inference cost and latency trade-offs of ResLPO compared with pointwise and listwise methods??
\end{itemize}

\begin{table*}[htbp]
\centering
\resizebox{\textwidth}{!}{%
\begin{tabular}{lllccccccccccccc}  
\toprule
\multirow{2}{*}{\textbf{Listwise}} &
\multirow{2}{*}{\textbf{Method}} &
\multirow{2}{*}{\textbf{Type}} &  
\multicolumn{3}{c}{\textbf{All\_Beauty}} &
\multicolumn{3}{c}{\textbf{Fashion}} &
\multicolumn{3}{c}{\textbf{Baby\_Products}} &
\multicolumn{3}{c}{\textbf{Software}} &
\textbf{Overall} \\
\cmidrule(lr){4-6} \cmidrule(lr){7-9} \cmidrule(lr){10-12} \cmidrule(lr){13-15} \cmidrule(lr){16-16}
 & & & N@1 & N@3 & N@10 & N@1 & N@3 & N@10 & N@1 & N@3 & N@10 & N@1 & N@3 & N@10 & NDCG \\
\midrule

\multirow{5}{*}{L=10}
 & BM25           & Pointwise & 0.509 & 0.649 & 0.851 & 0.523 & 0.670 & 0.860 & 0.511 & 0.642 & 0.811 & 0.504 & 0.630 & 0.842 & 0.853 \\
 & SFT & Pointwise & 0.672 & 0.790 & 0.916 & 0.778 & 0.875 & 0.946 & 0.700 & \textbf{0.817} & \textbf{0.927} & 0.748 & 0.832 & 0.884 & 0.918 \\
 & DPO            & Pairwise  & 0.467 & 0.611 & 0.824 & 0.527 & 0.647 & 0.872 & 0.490 & 0.623 & 0.852 & 0.519 & 0.654 & 0.861 & 0.853 \\
 & LIPO           & Listwise  & 0.630 & 0.743 & 0.890 & 0.668 & 0.770 & 0.903 & 0.658 & 0.778 & 0.913 & 0.718 & 0.786 & 0.911 & 0.904 \\
 & ResLPO (Ours)    & Hybrid    & \textbf{0.713} & \textbf{0.815} & \textbf{0.923} & \textbf{0.806} & \textbf{0.894} & \textbf{0.953} & \textbf{0.703} & 0.803 & 0.913 & \textbf{0.781} & \textbf{0.849} & \textbf{0.937} & \textbf{0.931} \\
\midrule

\multirow{5}{*}{L=20}
 & BM25           & Pointwise & 0.400 & 0.518 & 0.690 & 0.401 & 0.529 & 0.721 & 0.400 & 0.531 & 0.720 & 0.381 & 0.491 & 0.680 & 0.703 \\
 & SFT & Pointwise & 0.610 & 0.708 & 0.852 & 0.736 & 0.813 & 0.902 & 0.656 & 0.761 & 0.865 & 0.668 & 0.801 & 0.854 & 0.868 \\
 & DPO            & Pairwise  & 0.364 & 0.457 & 0.638 & 0.403 & 0.437 & 0.643 & 0.412 & 0.442 & 0.508 & 0.422 & 0.479 & 0.638 & 0.607 \\
 & LIPO           & Listwise  & 0.338 & 0.457 & 0.646 & 0.372 & 0.513 & 0.716 & 0.398 & 0.431 & 0.510 & 0.393 & 0.537 & 0.720 & 0.627 \\
 & ResLPO (Ours)    & Hybrid    & \textbf{0.702} & \textbf{0.776} & \textbf{0.877} & \textbf{0.761} & \textbf{0.847} & \textbf{0.919} & \textbf{0.697} & \textbf{0.778} & \textbf{0.881} & \textbf{0.675} & \textbf{0.768} & \textbf{0.859} & \textbf{0.878} \\
\midrule

\multirow{5}{*}{L=30}
 & BM25           & Pointwise & 0.362 & 0.452 & 0.513 & 0.342 & 0.446 & 0.619 & 0.345 & 0.457 & 0.637 & 0.306 & 0.408 & 0.590 & 0.594 \\
 & SFT & Pointwise & 0.572 & 0.713 & 0.815 & 0.640 & 0.778 & 0.870 & 0.633 & \textbf{0.723} & \textbf{0.828} & 0.629 & 0.739 & 0.821 & 0.845 \\
 & DPO            & Pairwise  & 0.324 & 0.388 & 0.576 & 0.349 & 0.420 & 0.597 & 0.352 & 0.389 & 0.572 & 0.372 & 0.402 & 0.606 & 0.588 \\
 & LIPO           & Listwise  & 0.297 & 0.393 & 0.561 & 0.348 & 0.420 & 0.647 & 0.301 & 0.393 & 0.573 & 0.311 & 0.403 & 0.582 & 0.612 \\
 & ResLPO (Ours)    & Hybrid    & \textbf{0.661} & \textbf{0.751} & \textbf{0.852} & \textbf{0.709} & \textbf{0.805} & \textbf{0.891} & \textbf{0.645} & 0.708 & 0.827 & \textbf{0.633} & \textbf{0.748} & \textbf{0.829} & \textbf{0.856} \\
\midrule

\multirow{5}{*}{L=50}
 & BM25           & Pointwise & 0.285 & 0.365 & 0.510 & 0.279 & 0.377 & 0.536 & 0.280 & 0.377 & 0.535 & 0.258 & 0.339 & 0.490 & 0.517 \\
 & SFT & Pointwise & 0.526 & \textbf{0.630} & 0.774 & 0.581 & 0.757 & 0.837 & \textbf{0.619} & \textbf{0.730} & \textbf{0.805} & 0.677 & 0.709 & 0.787 & 0.801 \\
 & DPO            & Pairwise  & 0.268 & 0.311 & 0.476 & 0.311 & 0.352 & 0.508 & 0.335 & 0.409 & 0.559 & 0.342 & 0.377 & 0.529 & 0.518 \\
 & LIPO           & Listwise  & - & - & - & - & - & - & - & - & - & - & - & - & - \\
 & ResLPO (Ours)    & Hybrid    & \textbf{0.573} & 0.617 & \textbf{0.791} & \textbf{0.644} & \textbf{0.776} & \textbf{0.860} & 0.615 & 0.713 & 0.799 & \textbf{0.706} & \textbf{0.726} & \textbf{0.811} & \textbf{0.809} \\
\bottomrule
\end{tabular}%
}
\caption{Performance comparison with different listwise length settings. SFT corresponds to ResLPO without residual contextualization. The best results in each block are highlighted in bold. ``--'' indicates failure to produce a valid full permutation (e.g., missing one or more candidates).}
\label{listwise_comparison}
\end{table*}

\subsection{Experimental Setup}
We use Mistral-7B-Instruct~\cite{jiang2023mistral7b} as the backbone LLM. Unless otherwise specified, we perform full-parameter fine-tuning rather than parameter-efficient adaptation (e.g., LoRA) in both Phase~1 (pointwise SFT) and Phase~2 (residual tuning). We compare ResLPO against a representative set of strong baselines, including Supervised Fine-Tuning (SFT), Direct Preference Optimization (DPO), Preference Ranking Optimization (PRO), and Listwise Preference Optimization (LIPO). 
For the DPO baseline, we construct the pairwise preference training set by sampling review pairs associated with the same product. To ensure a clear and robust preference signal, we only form pairs where the difference in their LLM-annotated utility scores is strictly greater than 1.0. 
To test robustness under varying context lengths, we adopt a dynamic list-size strategy: during training, the candidate list size $K$ is uniformly sampled up to 50 for each product, while at inference we evaluate under fixed list sizes $K \in \{10, 20, 30, 50\}$. We report NDCG at standard cutoffs, specifically NDCG@1, NDCG@3, and NDCG@10. All models are trained for 3 epochs with AdamW using a learning rate of $1\times 10^{-5}$.
Training is conducted on 8 NVIDIA B200 GPUs. For fair comparison, we fine-tune all backbone-based baselines with a per-device batch size of 1, resulting in an end-to-end training time of approximately 6 hours. The residual head in ResLPO is trained with a per-device batch size of 8 and converges in approximately 2 hours. Following Section~\ref{sec:dataset}, we use 10-fold cross-validation for training and evaluation.



\subsection{Results}
\subsubsection{Effectiveness Comparison (RQ1)} 
To assess the comparative effectiveness of ResLPO, we analyze the ranking performance across four distinct product domains under the standard listwise setting ($L=10$). As presented in Table. \ref{listwise_comparison}, ResLPO demonstrates consistent superiority over all baseline paradigms.
First, compared to the strong SFT (Pointwise) baseline, ResLPO achieves the highest NDCG scores across all categories. Specifically, in the \textit{All\_Beauty} domain, ResLPO improves NDCG@1 from 0.672 to 0.713 and NDCG@10 from 0.916 to 0.923. This trend holds for the \textit{Fashion}, \textit{Baby\_Products}, and \textit{Software} domains, culminating in an Overall NDCG of 0.931, surpassing the SFT baseline of 0.918. This validates our hypothesis that injecting global context via a residual head effectively corrects the calibration bias inherent in independent pointwise scoring.
Second, ResLPO significantly outperforms the Pairwise (DPO) baseline. We observe that DPO struggles to converge in this long-context ranking scenario, yielding an Overall NDCG of only 0.853. This suggests that pairwise objectives, which optimize local relative preferences, may be insufficient for capturing the global permutation structure required for high-utility review ranking, or they may suffer from optimization instability when scaling to dense lists.
Finally, while the standard Listwise (LIPO) method performs competitively at shorter list lengths (Overall NDCG 0.904 at $L=10$), it still lags behind ResLPO. ResLPO's hybrid architecture—combining the stability of pointwise semantic encoding with the context-awareness of the residual block—allows it to extract more precise ranking signals than the generative permutation likelihood objective used in LIPO.

We further observe that pointwise scoring is a strong and robust baseline in this setting. Across all list sizes, SFT (pointwise) consistently outperforms the pairwise DPO baseline, in line with the findings of \citet{gera2025justrank} that direct numeric scoring can be more effective than pairwise preference optimization for LLM ranking.
Finally, while LiPO is competitive at shorter lists, its performance degrades markedly as $L$ increases, and it fails at $L=50$ due to unstable generation (e.g., missing candidates in the produced permutation). This behavior is consistent with the long-context instability reported in \citet{liu2025lipo}: listwise generative ranking becomes increasingly brittle under long contexts, limiting its practical use to very small reranking sets (e.g., $L\leq 5$).

\subsubsection{Long-Context Robustness (RQ2)}

A critical challenge in LLM-based ranking is robustness to long candidate lists, where the lost-in-the-middle effect and other long-context artifacts can degrade performance as $K$ increases. As illustrated in Appendix~\ref{app2}, reviews in our benchmark can be lengthy; consequently, ranking a list of 50 reviews already corresponds to a realistic long-context setting. Scaling $K$ from 10 to 50 (Table~\ref{listwise_comparison}), we find that the generative listwise baseline LIPO deteriorates sharply at $K=20$ and $K=30$ and fails at $K=50$ (i.e., it cannot reliably output a complete permutation, often missing candidates), consistent with known long-context instability. In contrast, ResLPO remains stable across all lengths and is generally more robust than the pointwise SFT baseline at moderate list sizes, while at $K=50$ the gap narrows and each method has strengths in different domains. Overall, these results highlight a practical trade-off: pointwise scoring is inherently length-robust because it processes items independently, whereas ResLPO preserves listwise contextual benefits without the catastrophic failures that can arise in long-context generative listwise ranking.

\subsubsection{Generalization Across Domains (RQ3)}
To evaluate the transferability of the learned ranking policies, we conducted a cross-domain generalization experiment. We trained ResLPO on a single source domain and evaluated it zero-shot on the remaining three target domains under the standard setting ($K=10$). Table \ref{cross_domain_matrix} reports the NDCG@10 results, where diagonal elements represent in-domain performance and off-diagonal elements represent cross-domain transfer.

\begin{table}[htbp]
\centering
\resizebox{\columnwidth}{!}{%
\begin{tabular}{lcccc}
\toprule
\textbf{Train $\downarrow$ / Test $\rightarrow$} & \textbf{All\_Beauty} & \textbf{Fashion} & \textbf{Baby\_Products} & \textbf{Software} \\
\midrule
\textbf{All\_Beauty}      & \textbf{0.923} & 0.947 & 0.901 & 0.899 \\
\textbf{Fashion}          & 0.917 & \textbf{0.953} & 0.908 & 0.901 \\
\textbf{Baby\_Products}   & 0.903 & 0.939 & \textbf{0.913} & 0.872 \\
\textbf{Software}         & 0.898 & 0.902 & 0.897 & \textbf{0.937} \\
\bottomrule
\end{tabular}%
}
\caption{Cross-domain generalization performance (NDCG@10) of ResLPO. Rows indicate the source domain used for training, while columns indicate the target domain for evaluation. Diagonal elements (highlighted in bold) represent in-domain performance.}
\label{cross_domain_matrix}
\end{table}

The results reveal a remarkable degree of robustness. First, the performance gap between in-domain and cross-domain settings is minimal. For instance, the model trained on \textit{All\_Beauty} achieves an NDCG@10 of 0.947 when transferred to \textit{Fashion}, which is statistically comparable to the in-domain performance of the \textit{Fashion}-trained model (0.953). This suggests that ResLPO captures universal ranking signals—such as the correlation between review detail and utility—rather than overfitting to domain-specific product terminology.
Furthermore, ResLPO demonstrates that a robust listwise ranker can outperform domain-specific pointwise baselines even in a zero-shot setting. Referring back to the baselines in Table \ref{listwise_comparison}, the SFT model trained specifically on \textit{All\_Beauty} achieves an NDCG@10 of 0.916. Strikingly, the ResLPO model trained on \textit{Fashion} achieves a zero-shot score of 0.917 on \textit{All\_Beauty}, effectively matching the in-domain supervised baseline. Similarly, the \textit{Fashion}-trained model achieves 0.901 on \textit{Software}, surpassing the in-domain SFT performance for Software (0.884).
These findings confirm that the residual preference optimization objective learns generalized comparative reasoning skills that are highly transferable, reducing the need for extensive data annotation when deploying ranking models to new verticals.
We defer our detailed efficiency and scalability results (RQ4), including incremental latency under streaming updates, to Appendix~\ref{app4}.

\section{Conclusion}

ResLPO is a practical framework for long-context review ranking that balances effectiveness and efficiency through a residual design. Instead of performing expensive and unstable full listwise inference with an LLM over the entire candidate set, ResLPO first obtains strong pointwise scores for each review using a fine-tuned LLM, and then learns a list-conditioned residual term that adjusts these base scores using global list context—focusing the model capacity on correcting relative ordering errors rather than re-computing rankings from scratch. On a new benchmark derived from Amazon Reviews 2023 with LLM-based labels and human verification, ResLPO consistently outperforms strong pointwise, pairwise, and listwise baselines, while remaining stable as the candidate list grows to 50 reviews. Future work will extend this residual list-aware ranking architecture to other ranking scenarios (e.g., recommendation) and investigate how to integrate personalization signals and stronger scalable human evaluation.

\section*{Limitations}

Review utility is inherently subjective, and in many cases even expert annotators may find it difficult to reliably distinguish between two highly similar, high-quality reviews. This suggests that purely global helpfulness supervision may be insufficient for fine-grained tie-breaking, and incorporating user personalization signals is an important direction for future work. Second, while our human verification protocol based on iterative pairwise comparisons helps reduce noise and improves consistency, it is labor-intensive and does not scale well to large candidate sets, which limits the extent of human validation we can perform. Third, ResLPO is designed as a residual correction on top of a pointwise base scorer. When the base scorer is substantially miscalibrated or overly sensitive to prompt and style variations, the residual head may not fully compensate for these errors, particularly for rare, adversarial, or out-of-distribution reviews.


\bibliography{custom}
\bibliographystyle{acl_natbib}
\newpage

\appendix
\onecolumn
\section{LLM Annotation Prompt}
\label{app1}

\begin{appendixframe}[frametitle={Gemini-2.5-Pro Annotation Prompt}]
You are an e-commerce assistant at Amazon shop designed to output JSON format result, you are proficient in various languages.

\textbf{Background:}
A product review is a written assessment or evaluation of a product by a consumer who has used or experienced it. Product reviews typically include the consumer's opinions, feedback about various aspects of the product;

\textbf{Your task:}
Your task is to give a product review a ranking score from 1 to 10, which will be used to rank product reviews. The higher the ranking score, the higher the ranking of the review and easier to see it for consumers, thus helping consumers make a purchase decision.
You should give a score based on fully understanding the review content, based on the demision of Relevance, Quality, Usefulness, Content Richness and Objectivity.\\
1. Relevance: Ensure that the reviews are relevant to the product being reviewed or relevant to the shopping experience. Any off-topic reviews should be rated lower.\\
2. Quality: High-quality reviews contain detailed, well-structured opinions, and brief explanations. Avoid giving high scores to general reviews (such as "Good", "Great", or "Too bad"), repetitive reviews, completely capitalized review, reviews inclduding much exclamation and emoticon, purely emoji-based reviews, and reviews with 6 words or less.\\
3. Usefulness: give high scores to reviews that provide the most useful information to potential buyers higher. Useful reviews often include personal experiences, aspect-specific reviews (E.g 'Appearance', 'Arch support', 'Authenticity', 'Cleanability', 'Closure Type', 'Clothing Length', 'Clothing Styles', 'Clothing type', 'Color', 'Comfort', 'Concentration', 'Coverage', 'Design', 'Durability', 'Ease of Use', 'Ease of care', 'Ease of maintenance', 'Easy to remove', 'Effect on skin', 'Elasticity', 'Embellishment', 'Feature', 'Features', 'Finish', 'Fit', 'For Travel', 'Holiday', 'Ingredients', 'Layering', 'Leakage', 'Maneuverability', 'Material', 'Neckline', 'Occasion', 'Packaging', 'Pattern', 'Performance', 'Pockets', 'Portability', 'Purity', 'Quality', 'Quantity Per Pack', 'Scratch resista', 'Season', 'Shape', 'Sheer', 'Size', 'Skin tone match', 'Skin type', 'Smell', 'Smoothness', 'Staying power', 'Style', 'Texture', 'Theme', 'Transparency', 'Type', 'Value for money', 'Versatality', 'Versatility', 'Warmth', 'Wash', 'Waterproofness', 'Weaving Method', 'Weight', 'Wheels', 'Wind proof', 'Zipper', 'waterproofness') that help other buyers make informed decisions. The richer the aspects involved in the reviews, the higher the score should be.\\
4. Content Richness: Reviews should cover multiple aspects of the product, including strengths and weaknesses. High-score reviews should address customer concerns and provide informed information when customer make purchase decision.

You should finish the task strictly following below instructions:\\
1. The lower the score, the worse the quality of the review, and the higher the score, the better the quality of the review. Score accurately to 1 decimal place;\\
2. The score of a good review should be above 8 points. A good review should perfectly meet the requirements of high Relevance, high Quality, high Usefulness, high Content Richness and high Objectivity, and review content with 15 words or more.\\
3. The score for a moderate review should be between 5 and 8 points. Moderate review should meet the requirements of Relevance, Quality, Usefulness, Content Richness and Objectivity, but the writing quality of the review is not high enough, such as with some spelling errors, excessive use of Emoji, short content length with 10 words or less , etc.\\
4. Bad product review scores should be between 1 and 5 points. Usually refers to some reviews that do not meet the requirements of Relevance, Quality, Usefulness, Content Richness and Objectivity. Or contain some hateful and uncomfortable remarks.\\
5. If the review has no relevance to the product, the score should be lower;

Here are some examples for few-shot:\\
\textbf{Example1:}
    Product name: "Tower 28 Shineon Milky Lip Jelly in Cashew"
    Review: "i looovvveeedddd how smooth this product was. felt light and not sticky on the lips. pigment was there too but subtle which i loved"
    Output: {"score": 8.8, "explanation": "High-quality, useful, relevant, detailed content with \"not sticky\" aspects"}

\textbf{Example 2:}
    Product name: "Tower 28 SOS Daily Rescue Facial Spray 1oz"
    Review: "It made my skin feel so nice and refreshed and cleared up my acne so quick so 10/10 recommend!"
    Output: {"score": 8.5, "explanation": "High relevance, high quality, high usefulness, moderate content richness"}

\textbf{Example 3:}
    Product name: "Keep Up KanCan Flare Acid Wash Jeans"
    Review: "Too bad!"
    Output: {"score": 2.0, "explanation": "Content length is too short, generic review"}

\textbf{Example 4:}
    Product name: "Tower 28 SOS Daily Rescue Facial Spray 1oz"
    Review: "Way smaller than I thought it would be"
    Output: {"score": 6.5, "explanation": "high relevance, content length is too short, high usefulness, moderate content richness"}

\textbf{Example 5:}
    Product name: "LATTAFA HAYA EDP SPRAY Aroma Floral Fragrance Pack Perfume Scent Blend Scented Cosmetic Cologne"
    Review: "This smells so good and last all day! Its smell very similar to Viktor \& Rolf Good fortune which i absolutely love and the packaging is TOP TIER it gives luxury at a fraction of the price!!"
    Output: {"score": 9.2, "explanation": "High relevance, detailed quality, useful comparisons, rich content discussing smell and packaging."}

Product name: \{item\_title\} \\
Review: \{review\_text\} \\
Output:
\end{appendixframe}

\begin{multicols}{2}
\section{Visualization of Review Benchmark}
\label{app2}
We use the publicly available Amazon Reviews 2023 dataset. Since user-generated reviews may contain personally identifying information (PII) or offensive content, we rely on the dataset’s de-identification procedures, which remove fields such as user names/IDs and discard or mask obvious PII patterns (e.g., emails, phone numbers, addresses, and order numbers).
\end{multicols}

\begin{appendixframe}[frametitle={Sampled Reviews}]
\textbf{Example 1:}
\textbf{Item title:} Oral-B Vitality Dual Clean Rechargeable Electric Toothbrush\\
\textbf{Timestamp:} 1251995798000 \\
\textbf{Review Content:} I've had this brush for almost 2 months, and my teeth have never looked/felt better.  Because of its powerful scrubbing motions, this brush is doing most of the work for you.  No need for a death grip or vigorous brushing from you.  Simply gliding the brush over your teeth and gums is enough. I did notice minor bleeding and soreness on my gums within the first week of use, but that was because they weren't used to such a thorough cleaning.  No problems now.  The battery has been holding very well. I don't charge it all day because I like to conserve energy. It stays in top shape for at least 5-7 days before I have to charge again.  I only wished that it had a case or at least a brush head cover for traveling.  Pros: 1. Powerful brushing 2. Rechargeable 3. 2 minute timer  Cons: 1. No traveling cover/case  **Oct 2010 update** Just had my dental cleaning, and the hygienist told me she saw (and I quote) "superior brushing"! \\
\textbf{Score:} 9.9 \\
\textbf{Explanation:} This is an exceptionally high-quality and useful review. It is highly relevant, well-structured with a clear pros and cons list, and provides rich, detailed content based on two months of use. The review covers multiple specific aspects like performance ('powerful scrubbing'), battery life, and features (timer), while also noting a drawback (no travel case). The update from a dental professional adds significant credibility and usefulness, making it a near-perfect example of a helpful review.

\vspace{0.3cm}
\hrule
\vspace{0.3cm}

\textbf{Example 2:}
\textbf{Item title:} Oral-B Vitality Dual Clean Rechargeable Electric Toothbrush\\
\textbf{Timestamp:} 1268623912000 \\
\textbf{Review Content:} I bought this toothbrush because of all the statistics singing the praises of electric toothbrushes. I thought that this head looked particularly effective, so I placed my order. This is a great toothbrush. Some people have complained about the head size, but that's practically an Oral B trademark, and after a week or so you don't even notice. The toothbrush itself is pretty intense. It might hurt your gums the first time you use it. It is fast and powerful and it really gets the job done. The battery will last almost a week without charging, but after a few days you start to steadily lose power. I recommend keeping it charged most of the time. It's not particularly loud for a brush of its kind, but it does make some noise. My only complaint: the vibrations sometimes bother my lips and/or nose. It's not as noticeable after awhile, but it's a bit annoying at first. Still, that's not really the toothbrush's fault. It can't help being that intense. A final comment: Don't listen to some of the negative comments. A lot of them happened because the person didn't read the instructions before use. If used properly, the brush is a huge improvement over a manual. I think my dentist will appreciate it. Five stars. \\
\textbf{Score:} 9.82 \\
\textbf{Explanation:} Excellent review with high relevance, quality, and usefulness. The content is very rich, providing a balanced and detailed breakdown of the product's performance, battery life, noise level, and head size. It addresses both pros (powerful, effective) and cons (vibrations, initial sensitivity), making it extremely helpful for other customers making a purchase decision.

\vspace{0.3cm}
\hrule
\vspace{0.3cm}

\textbf{Example 3:}
\textbf{Item title:} Oral-B Vitality Dual Clean Rechargeable Electric Toothbrush\\
\textbf{Timestamp:} 1184828165000 \\
\textbf{Review Content:} The Oral B Vitality Dual Clean toothbrush performs like many of the \$100+ power brushes but at a fraction of the cost.  I started using mine about a month before my most recent dentist visit and noticed a definite improvement in my brushing results, both above and below the gumline.  Brushing with the Dual Clean takes a little getting used to.  First, it's a different technique than regular brushes - you simply glide the brush along your teeth and gums rather than scrubbing back and forth.  The fast pulsation of the head takes care of that for you.  Once you get use to this new technique, it makes for a very comfortable brushing experience;  however, I did experience some minor bleeding for the first few days of use.  Second, the power of the motor means that the vibration is intense until you get accustomed to it.  On the downside, this makes the handle rather bulky, although its rubberized grip makes it easy to manage.  Over the long run, the Dual Clean has proven to be well-constructed, having survived being packed away for several trips (battery life is good enough that you won't even need to bring the charger unless you plan on being away at least a week).  Maintenance is simple - both the handle and head are easy to keep clean by simply rinsing after each use.  Unfortunately, the initial value that this unit offers is diminished by the relatively high cost for replacement heads.  Overall, the Vitality Dual Clean does an excellent job of cleaning your entire mouth - my dentist said as much.  It's also dependable and very affordable, making it a great buy.  The sensation of a power toothbrush may not suit everyone's tastes, but at this price, it's easy to see for yourself. \\
\textbf{Score:} 9.8 \\
\textbf{Explanation:} Excellent review with extremely high relevance, quality, usefulness, and content richness. The reviewer provides a comprehensive, well-structured analysis covering numerous aspects like performance, value for money, ease of use, durability, battery life, and both pros and cons (e.g., expensive replacement heads). This detailed and balanced personal experience is exceptionally helpful for potential buyers.
\end{appendixframe}

\begin{multicols}{2}
\section{Additional Baselines: Encoder-based and Zero-shot LLM Rankers}
\label{app:additional_baselines}

To further strengthen our evaluation and provide a more comprehensive comparison across different ranking paradigms, we include representative encoder-based rerankers and LLM zero-shot rankers as additional baselines. Specifically, we evaluate bge\_reranker-v2-m3 \cite{chen2024bge} as a state-of-the-art encoder model and rank\_vicuna\_7b \cite{pradeep2023rankvicuna} as a representative zero-shot LLM ranker. 

Table \ref{tab:additional_baselines} presents the NDCG results of these models on the \textit{All\_Beauty} domain across varying list lengths. The results demonstrate that while these baselines perform reasonably well at shorter list lengths (e.g., $L=10$), their performance degrades significantly as the context length increases. Notably, the zero-shot LLM ranker fails to produce valid permutations for longer lists ($L=30$ and $L=50$), further highlighting the long-context robustness of our proposed ResLPO framework.

\end{multicols}

\begin{table}[htbp]
\centering
\small
\begin{tabular}{lllccc}
\toprule
\textbf{Listwise} & \textbf{Method} & \textbf{Type} & \textbf{N@1} & \textbf{N@3} & \textbf{N@10} \\
\midrule
\multirow{2}{*}{L=10} 
 & bge\_reranker & encoder & 0.570 & 0.702 & 0.874 \\
 & rank\_vicuna & LLM zero shot & 0.678 & 0.739 & 0.885 \\
\midrule
\multirow{2}{*}{L=20} 
 & bge\_reranker & encoder & 0.467 & 0.580 & 0.756 \\
 & rank\_vicuna & LLM zero shot & 0.457 & 0.494 & 0.577 \\
\midrule
\multirow{2}{*}{L=30} 
 & bge\_reranker & encoder & 0.421 & 0.523 & 0.684 \\
 & rank\_vicuna & LLM zero shot & -- & -- & -- \\
\midrule
\multirow{2}{*}{L=50} 
 & bge\_reranker & encoder & 0.414 & 0.488 & 0.625 \\
 & rank\_vicuna & LLM zero shot & -- & -- & -- \\
\bottomrule
\end{tabular}
\caption{Performance of additional encoder-based and zero-shot LLM baselines on the \textit{All\_Beauty} domain. ``--'' indicates failure to produce a valid full permutation for the given list length.}
\label{tab:additional_baselines}
\end{table}

\begin{multicols}{2}
\section{Human Evaluation Dimensions}
\label{app3}

We conduct human evaluation under two complementary protocols: (i) a listwise setting that asks annotators to score and rank the top-50 reviews for each product, and (ii) a pairwise setting that asks annotators to compare two reviews at a time. Both protocols share a common set of core dimensions (quality, relevance, emotion, and expression), while the listwise setting additionally produces a global ranking and a tie-breaking preference aligned with purchase appeal and brand value. Table~\ref{tab:human_eval_dims} summarizes the annotation fields and criteria. The three annotators were recruited internally; participation was voluntary and they were compensated at a standard hourly rate. We provided written instructions and asked annotators to stop if they encountered uncomfortable content.
\end{multicols}

\begin{table}[t]
\centering
\footnotesize
\setlength{\tabcolsep}{4pt}
\begin{tabular}{p{2.0cm} p{6.0cm} p{3.1cm} p{3.2cm}}
\toprule
\textbf{Protocol} & \textbf{Core Rating Dimensions} & \textbf{Auxiliary Checklist (Yes/No)} & \textbf{Final Outputs} \\
\midrule
\textbf{Listwise} 
&
\textbf{Input:} \texttt{review\_content}. \newline
\textbf{Ratings (0--10 each):} (1) Quality of review, (2) Relevance between review and product, (3) Emotion of review, (4) Expression/clarity of review. \newline
\textbf{Total score:} sum of the four ratings.
&
(1) Includes multi-dimensional product info (e.g., color/size/style)? \newline
(2) Includes sufficient details? \newline
(3) Compares with similar products / shows competitiveness? \newline
(4) Objective / true / credible? \newline
(5) Content related to the product? \newline
(6) Positive review? \newline
(7) Increases desire to purchase? \newline
(8) Expression clear and logical?
&
\textbf{Ranking (1--50)} based on total score. \newline
\textbf{Tie-breaker:} if totals tie, prefer the review that is more appealing for purchase and better reflects product/brand value.
\\
\midrule
\textbf{Pairwise}
&
\textbf{Input:} \texttt{review\_content\_v1}, \texttt{review\_content\_v2}. \newline
\textbf{Ratings (0--5 each, per review):} (1) Quality, (2) Relevance, (3) Emotion, (4) Expression/clarity. \newline
\textbf{Total score:} sum of the four ratings (computed per review).
&
(1) Includes multi-dimensional product info? \newline
(2) Includes sufficient details? \newline
(3) Compares with similar products / competitiveness? \newline
(4) Objective / true / credible? \newline
(5) Content related to the product? \newline
(6) Increases desire to purchase? \newline
(7) Expression clear and logical?
&
\textbf{Winner:} \texttt{review\_v1} or \texttt{review\_v2}. \newline
\textbf{Sentiment labels:} \texttt{sentiment\_v1} (1--5), \texttt{sentiment\_v2} (1--5).
\\
\bottomrule
\end{tabular}
\caption{Human evaluation dimensions and outputs for listwise (Top-50) and pairwise protocols. Both share four core dimensions; listwise additionally yields a global ranking with a purchase/brand-oriented tie-break rule.}
\label{tab:human_eval_dims}
\end{table}

\begin{multicols}{2}
\section{Efficiency and Scalability}
\label{app4}
We measure incremental inference latency under a streaming update scenario: each time a new review is added, the system computes the necessary scores to integrate this review into ranking. We report the \emph{mean end-to-end latency} averaged over 20 runs (after 5 warm-up runs). All methods use the same backbone (Mistral-7B-Instruct) and the same decoding/tokenization stack; token generation speed is reported to control for hardware/runtime effects.
Table~\ref{tab:efficiency} summarizes the average per-new-review latency. For pointwise SFT, adding one review requires a single pointwise forward/generation pass. ResLPO adds a lightweight representation-level residual step on top of the pointwise scorer, resulting in a modest overhead relative to SFT. In contrast, LiPO (a generative listwise ranker) incurs substantially higher latency, consistent with token-level listwise processing and the need to generate/verify a full permutation as the candidate list grows.
\end{multicols}

\begin{table}[h]
\centering
\small
\begin{tabular}{lccc}
\toprule
\textbf{Method} & \textbf{Granularity} & \textbf{Mean latency} $\downarrow$ & \textbf{Token speed} \\
\midrule
SFT  & per-review (pointwise) & 1.4377s & 32.00 tok/s \\
LiPO & per-list (generative listwise) & 14.512s & 31.78 tok/s \\
ResLPO & per-review + residual head & 1.8377s & 32.21 tok/s \\
\bottomrule
\end{tabular}
\caption{Incremental inference cost of adding a single new review to a product’s review list. Token processing speed is identical across methods, so latency differences mainly reflect algorithmic overhead rather than hardware or runtime variability.}
\label{tab:efficiency}
\end{table}

\begin{multicols}{2}
For readability, we also compute effective throughput as the reciprocal of latency (lists/sec for LiPO; reviews/sec for SFT/ResLPO):
\[
\text{throughput} \approx \frac{1}{\text{latency}}.
\]
This yields $\sim 0.70$ reviews/sec for SFT, $\sim 0.54$ reviews/sec for ResLPO, and $\sim 0.069$ lists/sec for LiPO under our setup.
\end{multicols}

\end{document}